# Symbolic Simulation-Checking of Dense-Time Systems [*]

Farn Wang

Dept. of Electrical Engineering, National Taiwan University
1, Sec. 4, Roosevelt Rd., Taipei, Taiwan 106, ROC;
+886-2-33663602; FAX:+886-2-23671909;
farn@cc.ee.ntu.edu.tw; http://cc.ee.ntu.edu.tw/~farn

Library **RED** 7.0 is available at http://cc.ee.ntu.edu.tw/~val

**Abstract.** Intuitively, an *(implementation) automata is simulated by* a *(specification) automata* if every externally observable transition by the implementation automata can also be made by the specification automata. In this work, we present a symbolic algorithm for the simulation-checking of timed automatas. We first present a simulation-checking procedure that operates on state spaces, representable with convex polyhedra, of timed automatas. We then present techniques to represent those intermediate result convex polyhedra with zones and make the procedure an algorithm. We then discuss how to handle Zeno states in the implementation automata. Finally, we have endeavored to realize the algorithm and report the performance of our algorithm in the experiment.

## 1 Introduction

In the last two decades, we have witnessed significant progress in both theory and applications of the formal verification of real-time systems. Especially, the technology of dense-time system *model-checking* [1] has been well-received by the academia, realized with many tools [8,19,22], and used for the verification of several industrial projects [6]. With model-checking, we represent the *implementation* as an *automata* (*state-transition diagram* or *table*) and the *specification* as a temporal logic formula and want to check whether the implementation satisfies the specification. However, we have to admit that the promise of the model-checking technology of real-time systems has not been fulfilled yet. One reason is that engineers are not trained in writing formulas in temporal logics, like *TCTL (Timed Computational-Tree Logic)* [1]. In many applications, engineers may also envision their *specifications* as automatas. With model-checking, it is usually difficult to completely and correctly represent a specification automata as a set of logic formulas. There are two common frameworks for checking implementation automatas against specification ones. The first is the *language inclusion problem* which checks if all runs of an implementation are also those of the specification. It was proved in [3] that when both the implementation and specification are represented as *timed automatas (TA)* [3], the language inclusion problem is undecidable. The second is called the *simulation-checking problem* that intuitively checks if every transition that can be performed by the implementation can also be matched by the specification at the same instant. (Formal definition in page 4.) It has been proved that the simulation checking problem of TAs is in EXPTIME [18]. However, the algorithms in the literature are either based on region graph analysis [1] of the product of the implementation and the specification TAs [18] or based on time-abstraction [13], which does not preserve the timing properties. In this work, we have the following contributions.

[*] The work is partially supported by NSC, Taiwan, ROC under grants NSC 94-2213-E-002-085 and NSC 94-2213-E-002-092.

- *A symbolic procedure that checks for the simulation relation between dense-time systems.* The procedure handles convex polyhedra in dense spaces of variables and straightforwardly falls in the realm of state-representation manipulation of *linear hybrid automatas (LHA)* [2]. Thus the procedure is good for the simulation-checking of both LHAs and TAs.
- *Techniques to implement the above-mentioned procedure with zone-technology.* In general, the manipulation of convex polyhedra can be very complex and the verification problem of LHAs is undecidable. As a special subclass of LHAs, the state-spaces of TAs can be efficiently represented and manipulated with *zones*[1] [14]. In section 6, we present techniques to represent the intermediate results of the above-mentioned procedure with zones. The techniques effectively make the procedure a symbolic algorithm.
- *A technique to handle Zeno states in simulation checking.* A *Zeno state* is one from which no computation yields divergent computation time. Intuitively, if the implementation TA can transit to a Zeno state, such a transition need not be matched by the specification TA and should not affect the answer to the simulation-checking problem. In section 7, we present a lemma that helps us extending our simulation-checking algorithm to handle Zeno states in the implementation TAs.

We have realized our algorithm and techniques with TCTL model-checker **RED**, version 7. Following is our presentation plan. Section 2 discusses related work. Section 3 defines our modeling language, TAs extended with event notations. Section 4 gives the definition and symbolic representation of the simulation relation between two TAs. Section 5 derives a symbolic procedure out of the definition of simulation relation. Section 6 discusses how to implement the procedure in section 5 with zones. Section 7 discusses how to filter out false negation from Zeno states of the implementation TA. Section 8 reports our program and the experiment. Section 9 is the conclusion.

---

[1] A *zone* is a conjunction of atomic propositions and constraints like either $x - y \leq c$ or $x - y < c$ where $x, y$ are either zero or clocks and $c$ is an integer constant.

## 2  Related work

Cerans showed that the bisimulation-checking problem of timed processes is decidable [9]. Taşıran et al showed that the simulation-checking problem of TAs is in EXPTIME [18]. They also proposed an algorithm to check whether a location homomorphism between an implementation TA and a specification TA preserves timed behaviors . However, there is no general strategy to efficiently construct such homomorphisms. In comparison, our approach is purely algorithmic and automatic.

Henzinger et al presented an algorithm that computes the time-abstract simulation that does not preserve timed properties [13].

Nakata also discussed how to do symbolic bisimulation checking with integer-time labelled transition systems [16]. Beyer has implemented a refinement-checking algorithm for TAs with integer-time semantics [7]. In comparison, our algorithm is for dense-time semantics.

Lin and Wang presented a complete and sound proof system for the equivalence of TAs with dense-time semantics [15]. Usually, the proofs may need human guidance.

Aceto et al constructed a modal logic formula that completely characterizes a TA [4]. Thus the simulation checking problem can be reduced to the model-checking problem. However, the formula they constructed is not for TAs with timed invariance constraints and does not handle the effect of Zeno states. In a sense, our formula (A) in definition 4 is also such a formula for TA with timed invariance constraints. Specifically, their characteristic formula is also of greatest fixpoint in nature and falls in the realm of LHA verification. It is not clear whether they can efficiently evaluate such characteristic formulas. In contrast, we have proposed and implemented the simulation-checking algorithm with zone technology.

## 3  A modeling language of dense-time systems

We need the following notations for convenience of presentation. Given a set $P$ of atomic propositions and a set $X$ of clocks, we use $B(P, X)$ as the set of all Boolean combinations of atoms of the forms $p$ and $x \sim c$, where $p \in P$, $x \in X \cup \{0\}$, '$\sim$' is one of $\leq, <, =, >, \geq$, and $c$ is an integer constant. An element in $B(P, X)$ is called a *state predicate*.

A *valuation* of a set $Y$ *(domain)* is a mapping from $Y$ to a *codomain*. A *partial valuation* is undefined for some elements in the domain. When it is not said specifically, a valuation means a *total valuation* that assigns a value to every element in the domain. A valuation $\nu$ *satisfies* a state-predicate $\eta$, in symbols $\nu \models \eta$, if the state-predicate evaluates to true when all its variables are interpreted according to $\nu$. Given two (partial) valuations $\Pi$ and $\Pi'$ on domain $Y$, $\Pi\Pi'$ is a new valuation defined in the following way. For all $y \in Y$,

- if $\Pi'(y)$ is defined, $\Pi\Pi'(y) = \Pi'(y)$;

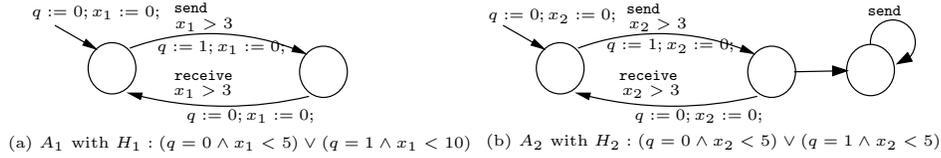

(a) $A_1$ with $H_1 : (q = 0 \wedge x_1 < 5) \vee (q = 1 \wedge x_1 < 10)$  (b) $A_2$ with $H_2 : (q = 0 \wedge x_2 < 5) \vee (q = 1 \wedge x_2 < 5)$

**Fig. 1.** Two example TEAs

- else if $\Pi'(y)$ is undefined and $\Pi(y)$ is defined, $\Pi\Pi'(y) = \Pi(y)$;
- else $\Pi\Pi'(y)$ is undefined.

We use the following extension of TA [3] for the modeling language in this work.

**Definition 1. <u>timed event-automata (TEA)</u>** A *timed event-automata (TEA) A* is a tuple $\langle \Sigma, X, G, L, I, H, E, \epsilon, \tau, \pi \rangle$ with the following restrictions. $\Sigma$ is a finite set of event names. $X$ is a finite set of clocks. $G$ is a finite set of global atomic propositions. $G$ and $\Sigma$ represent the set of external observables of a TEA. $L$ is a finite set of local atomic propositions such that $G \cap L = \emptyset$. $I \in B(G \cup L, X)$ is the initial condition. $H \in B(G \cup L, X)$ is the invariance condition. $E$ is the finite set of transitions. $\epsilon : E \mapsto 2^\Sigma$ labels each rule with a set of input/output events. $\tau : E \mapsto B(G \cup L, X)$ defines the triggering condition of each rule execution. For each $e \in E$, $\pi(e)$ is a partial valuation from $X$ to $\{0\}$ and from $G \cup L$ to $\{true, false\}$ that defines the assignments to clocks and proposition variables of each rule execution. If $\pi(e)(y)$ is undefined, it means variable $y$ stays unchanged in the transition.

For convenience, we assume that for every TEA, there is a null transition $\bot$ such that $\epsilon(\bot) = \emptyset$, $\tau(\bot) = true$, and $\pi(\bot)$ is undefined on everything. ■

*Example 1.* We have the transition diagrams of two example TEAs in figure 1. They share events send and receive and global proposition $q$. They respectively have local clocks $x_1$ and $x_2$. $A_1$ has two transitions while $A_2$ has four. We stack the events, triggering conditions, and the assignments made at each transition. The initial conditions are labeled by the arcs without a source. ■

Let $\mathbb{N}$ be the set of non-negative integers and $\mathbb{R}^{\geq 0}$ the set of non-negative reals.

**Definition 2. <u>States of a TEA</u>** A *state* of TEA $A$ is a total valuation from $X$ to $\mathbb{R}^{\geq 0}$ and $G \cup L$ to $\{true, false\}$. Let $V_A$ denote the set of states of $A$. For any state $\nu$ and $\delta \in \mathbb{R}^{\geq 0}$, $\nu + \delta$ is a valuation identical to $\nu$ except that for every $x \in X$, $\nu(x) + \delta = (\nu + \delta)(x)$. ■

Given two states $\nu, \nu'$ and a transition $e$, we say $A$ *transits with $e$ from $\nu$ to $\nu'$*, in symbols $\nu \xrightarrow{e} \nu'$, if $\nu \models \tau(e)$, $\nu\pi(e) = \nu'$, and $\nu' \models H$. Based on the above-presented notions, we are ready to define linear computations of TEAs.

**Definition 3. <u>runs</u>** Given a TEA $A = \langle \Sigma, X, G, L, I, H, E, \epsilon, \tau, \pi \rangle$, a *run* is an infinite computation of $A$ along which time diverges. Formally speaking, a run is an infinite sequence of state-time pairs $(\nu_0, t_0)(\nu_1, t_1) \ldots (\nu_k, t_k) \ldots \ldots$ such that

- $t_0 t_1 \ldots t_k \ldots \ldots$ is a monotonically increasing divergent real-number sequence, i.e., $\forall c \in \mathbb{N}, \exists k > 1, t_k > c$; and
- for all $k \geq 0$, for all $\delta \in [0, t_{k+1} - t_k]$, $\nu_k + \delta \models H$; and
- for all $k \geq 0$, there is an $e \in E$ such that $\nu_k + t_{k+1} - t_k \xrightarrow{e} \nu_{k+1}$. ∎

## 4 Implementation, simulation, refinement, and equivalence

Suppose we are given two TEAs $A_1$ and $A_2$ with $A_i = \langle \Sigma, X_i, G, L_i, I_i, H_i, E_i, \epsilon_i, \tau_i, \pi_i \rangle$, $1 \leq i \leq 2$. Intuitively, $A_1$ *implements* (or *refines*, or *is simulated by*) $A_2$ if we can map every reachable state $\nu_1$ of $A_1$ to a reachable state $\nu_2$ of $A_2$ such that every externally observable that $A_1$ can do at a specific time from $\nu_1$, $A_2$ can also do it at the same instant from $\nu_2$. If $A_2$ can always direct its runs so that no difference in external behaviors of $A_1$ and $A_1$ will ever be observed, then we say $A_1$ *implements* $A_2$. This is formalized with the following definition.

We need the following notations for the convenience of discussion. Two transitions $e_1 \in E_1$ and $e_2 \in E_2$ are *compatible* if $\epsilon_1(e_1) = \epsilon_2(e_2)$ and $\forall p \in G(\pi_1(p) = \pi_2(p))$. Given an $e_1 \in E_1$, we let $E_2^{(e_1)} = \{e_2 \mid e_2 \in E_2; e_2 \text{ is compatible with } e_1.\}$. Given $\alpha \in V_{A_1}$, $\beta_1, \beta_2 \in V_{A_2}$, $\delta \in \mathbb{R}^{\geq 0}$, and $Q \subseteq V_{A_1} \times V_{A_2}$, $stutter_2(\alpha, \beta_1, \delta, \beta_2, Q)$ is true iff

- $A_2$ can go from $\beta_1$ to $\beta_2$ through a series of time-progression steps and discrete transition steps in $E_2^{(\perp)}$;
- the finite run from $\beta_1$ to $\beta_2$ is $\delta$ time units long; and
- for any $0 \leq \delta' \leq \delta$ and state $\beta' \in V_{A_2}$ that is $\delta'$ time units away from $\beta_1$ in the run, $(\alpha + \delta', \beta') \in Q$ also.

Formally speaking, $stutter_2(\alpha, \beta_1, \delta, \beta_2, Q)$ is true iff there is a finite run $(\bar{\nu}_0, t_0)(\bar{\nu}_1, t_1) \ldots (\bar{\nu}_k, t_k)$ of $A_2$ and $t_{k+1} \geq t_k$ such that

- $t_{k+1} - t_0 = \delta \land \beta_1 = \bar{\nu}_0 \land \beta_2 = \bar{\nu}_k + t_{k+1} - t_k$; and
- $\forall 0 \leq h \leq k \forall \delta \in [t_{h+1} - t_h], (\alpha_1 + t_h - t_0 + \delta, \bar{\nu}_h + \delta) \in Q$; and
- $\forall 0 \leq h < k \exists \bar{e}_{h+1} \in E_2^{(\perp)}, \bar{\nu}_h + t_{h+1} - t_h \xrightarrow{\bar{e}_{h+1}} \bar{\nu}_{h+1}$.

**Definition 4. <u>Implementation, simulation, and equivalence</u>** Suppose we are given two TEAs $A_1$ and $A_2$ such that $A_i = \langle \Sigma, X_i, G, L_i, I_i, H_i, E_i, \epsilon_i, \tau_i, \pi_i \rangle$, $1 \leq i \leq 2$, $L_1 \cap L_2 = \emptyset$, and $X_1 \cap X_2 = \emptyset$. A *simulation relation* $Q$ from $A_1$ to $A_2$ is a binary relation from $V_{A_1}$ to $V_{A_2}$ with the following restriction. For every $(\nu_1, \nu_2) \in Q$, $\nu_1$ and $\nu_2$ agree on interpretation of variables in $G$ and for every $\delta \in \mathbb{R}^{\geq 0}$ and transition $e_1$ of $E_1$ such that for all $\hat{\delta} \in [0, \delta]$, $\nu_1 + \hat{\delta} \models H_1$ and $\nu_1 + \delta \xrightarrow{e_1} \nu_1'$, there are $\nu_2' \in V_{A_2}$ and $e_2 \in E_2^{(e_1)}$ such that $stutter_2(\nu_1, \nu_2, \delta, \nu_2', Q)$, $\nu_2' \xrightarrow{e_2} \nu_2' \pi_2(e_2)$, and $(\nu_1', \nu_2' \pi_2(e_2)) \in Q$. Or equivalently, $(\nu_1, \nu_2) \notin Q$ if there is $e_1 \in E_1$ satisfying formula (A) in the following.

$$\exists \delta \in \mathbb{R}^{\geq 0} \left( \begin{array}{c} \left( \begin{array}{c} \nu_1 + \delta \models \tau_1(e_1) \land (\nu_1 + \delta) \pi_1(e_1) \models H_1 \\ \land \neg \exists 0 \leq \delta' \leq \delta (\nu_1 + \delta' \models \neg H_1) \end{array} \right) \\ \land \neg \exists \nu_2' \in V_{A_2} \left( \begin{array}{c} stutter(\nu_1, \nu_2, \delta, \nu_2', Q) \\ \land \exists e_2 \in E_2^{(e_1)} \left( \begin{array}{c} \nu_2' \xrightarrow{e_2} \nu_2' \pi_2(e_2) \\ \land ((\nu_1 + \delta) \pi_1(e_1), \nu_2' \pi_2(e_2)) \in Q \end{array} \right) \end{array} \right) \end{array} \right) \ldots \text{(A)}$$

Note that formula (A) has the following structure: $\exists \delta \in \mathbb{R}^{\geq 0}(B \wedge \neg C)$. Formula $(B)$ says that $A_1$ can do transition $e_1$ at time $\delta$ from $\nu_1$. Formula $(C)$ says that $A_2$ can match $e_1$ with $e_2$ at the same time from $\nu_2$ through a run with transitions internal to $A_2$. The quantification scope of $\delta$ contains both $(B)$ and $(C)$ to make sure $e_1$ and $e_2$ are matched at the same instant.

Given simulation relation $Q$ from $A_1$ to $A_2$, we denote $A_1 \preceq_Q A_2$ if for every state $\nu_1 \models I_1$, there is a $\nu_2 \models I_2$ with $(\nu_1, \nu_2) \in Q$. If $\exists Q(A_1 \preceq_Q A_2)$, we say $A_1$ *implements* (or *refines*, or *is simulated by*) $A_2$, in symbols $A_1 \preceq A_2$. Two TEAs $A_1$ and $A_2$ *are equivalent* (i.e. *bisimulate*), in symbols $A_1 \equiv A_2$, if $A_1 \preceq A_2$ and $A_2 \preceq A_1$. ∎

*Example 2.* In example 1, $A_1$ is not simulated by $A_2$ since $A_1$ can make a transition after 5 time units in state $q = 1$ while $A_2$ cannot. ∎

In a TEA, there could be some computation that does not yield divergent computation time. Specifically, a *Zeno computation* is an infinite run $(\nu_0, t_0) \ldots (\nu_k, t_k) \ldots$ such that its time sequence $t_0 \ldots t_k \ldots$ converges to a finite value. Zeno computations are counter-intuitive. A *Zeno state* is a state that only starts Zeno computations. The problem with Zeno states in simulation-checking is that $A_1$ may stay in Zeno states that are not matched by any specified state of $A_2$. Intuitively, we want to check that $A_1$ implements $A_2$ from all non-Zeno states. We present the following definition to make this clear.

**Definition 5.** *Non-Zeno implementation and equivalence* Let $NZ_1$ be a representation of the non-Zeno states in the reachable state-space of $A_1$. An *NZ-simulation* $Q$ from $A_1$ to $A_2$ is a binary relation from $V_{A_1}$ to $V_{A_2}$ that satisfies the following requirement. For every $(\nu_1, \nu_2) \in Q$, if $A_1$ can do a transition $e_1$ at $\delta$ time units from $\nu_1$ "**to a non-ZENO state**" $\nu_1'$, then $A_2$ can also do a transition $e_2 \in E_2^{(e_1)}$ after a finite run of $\delta$ time units long with transitions internal to $A_2$ to a state $\nu_2'$ such that $(\nu_1', \nu_2') \in Q$. The only difference from definition 4 is th the bold-face phrase in the last sentence. The formal definition is left to appendix A due to page-limit. $A_1$ *NZ-implements* (or *NZ-refines*, or *is NZ-simulated by*) $A_2$, in symbols $A_1 \preceq^{NZ} A_2$, if there is a non-Zeno simulation relation $Q$ from $A_1$ to $A_2$ such that for every non-Zeno state $\nu_1 \models I_1 \wedge NZ_1$, there is a $\nu_2 \models I_2$ such that $(\nu_1, \nu_2) \in Q$. Two TEAs $A_1$ and $A_2$ *are NZ-equivalent* (i.e. *NZ-bisimulate*), in symbols $A_1 \equiv^{NZ} A_2$, if $A_1 \preceq^{NZ} A_2$ and $A_2 \preceq^{NZ} A_1$. ∎

*Example 3.* In example 1, $A_2$ is not simulated by $A_1$ since $A_2$ can yield infinite sequences of the `send` events while $A_1$ cannot. Such sequences are from Zeno states with $x_2$'s value converging to 5. In fact, $A_2$ is NZ-simulated by $A_1$. ∎

## 5 Symbolic procedure for simulation-checking

Formula (A) leads to a greatest fixpoint procedure for calculating a simulation relation from $A_1$ to $A_2$ if any. The idea is to first compute an initial image of $Q$ and then iteratively delete state pairs from $Q$ with formula (A) until a fixpoint is reached. Please be reminded that formula (A) has the following structure: $\exists \delta \in \mathbb{R}^{\geq 0}(B \wedge \neg C)$. In the following, we first present a scheme for the symbolic representation of $Q$. Then we discuss how to construct formulas (B) and (C) respectively out of $Q$.

### 5.1 Symbolic representation of $Q$

We define *linear hybrid predicates (LH-predicates)* to represent convex polyhedra in dense space. An LH-predicate is a Boolean combination whose atoms are either
- atomic proposition, like $p$, or
- linear constraints like $\sum_{1 \leq i \leq n} c_i x_i \sim d$ where $x_1, \ldots, x_n$ are clock variables, $c_1, \ldots, c_n, d$ are integer constants, and '$\sim$' $\in \{\leq, <\}$.

Let $C_{(A_1, A_2)}$ be the biggest constants used in $A_1$ and $A_2$. An LH-predicate is called a *zone-predicate* if its linear constraints are like $x_1 - x_2 \sim d$ where $x_1, x_2$ are either zeros or clock variables and $d \in \mathbb{N} \cap [0, C_{(A_1, A_2)}]$. Clearly, zone-predicates are special cases of LH-predicates. In practice, such zone-predicates can be implemented with DBMs [10] or CRDs [19] while LH-predicates can be with convex polyhedra [12] or HRDs [20]. Given a $(\nu_1, \nu_2) \in Q$, we require in definition 4 that $\nu_1$ and $\nu_2$ agree on the interpretation of variables in $G$. Thus if we have LH-predicates $\eta_1$ and $\eta_2$ for $\nu_1$ and $\nu_2$ respectively, we can use $\eta_1 \wedge \eta_2$ to represent pairs like $(\nu_1, \nu_2)$ in $Q$.

*Example 4.* Given $G = \{a\}, X_1 = \{x_1\}, L_1 = \{b_1\}, X_2 = \{x_2\}, L_2 = \{b_2\}$, we may have the following LH-predicate (also zone-predicate) $(f_Q)$ for $Q$.

$$\begin{array}{l}(a \wedge b_1 \wedge \neg b_2 \wedge 0 \leq x_1 \wedge 3 < x_2 \leq 5 \wedge x_2 - x_1 \leq 5) \\ \vee (\neg a \wedge 2 \leq x_1 < 9 \wedge 1 < x_2 \wedge x_1 - x_2 < 8)\end{array} \quad \ldots\ldots\ldots\ldots (f_Q)$$

∎

In the following, we propose procedures that manipulate LH-predicates to represent pairs like $(\nu_1, \nu_2)$.

### 5.2 Basic building blocks

One fundamental procedure is *Fourier-Motzkin elimination* [11]. Suppose we have a formula $F$ of variables in set $Y$. In this work, Fourier-Motzkin elimination constructs $\exists y(F)$ as a formula without $y \in Y$. For LH-predicates and zone-predicates, efficient implementation of Fourier-Motzkin elimination has been discussed in [14, 19, 20].

*Example 5.* For formula $(f_Q)$ in example 4,

$$\exists a(f_Q) \equiv \begin{array}{l}(b_1 \wedge \neg b_2 \wedge 0 \leq x_1 \wedge 3 < x_2 \leq 5 \wedge x_2 - x_1 \leq 5) \\ \vee (2 \leq x_1 < 9 \wedge 1 < x_2 \wedge x_1 - x_2 < 8)\end{array}$$

And $\exists x_1(f_Q) \equiv (a \wedge b_1 \wedge \neg b_2 \wedge 3 < x_2 \leq 5) \vee (\neg a \wedge 1 < x_2 < 6)$. ∎

We also need the following procedures to present our procedure. Given a symbolic representation $\eta$ of $\nu$, the symbolic representation of $\nu + \delta$ can be obtained by replacing each clock $x$ in $\eta$ with $x + \delta$ [14].

*Example 6.* For the $(f_Q)$ in example 4, $f_Q + \delta$ is the following.

$$(a \wedge b_1 \wedge \neg b_2 \wedge 0 \leq x_1 + \delta \wedge 3 < x_2 + \delta \leq 5 \wedge x_2 - x_1 \leq 5)$$
$$\vee (\neg a \wedge 2 \leq x_1 + \delta < 9 \wedge 1 < x_2 + \delta \wedge x_1 - x_2 < 8)$$

Note that for constraints like $x_1 - x_2 < 8$, the positive and negative $\delta$ respectively from $x_1$ and $x_2$ cancel with each other. ∎

Let $\texttt{path}(\eta_1, \delta)$ be the constraint that from now to $\delta$ time units in the future, $\eta_1$ is true, i.e., $\texttt{path}(\eta_1, \delta) \equiv \neg \exists \delta'((\neg \eta_1) + \delta' \wedge 0 \leq \delta' \wedge \delta' \leq \delta)$. We can then define $\texttt{Tbck}(\eta', \eta)$ that computes the space representation of states

- from which we can go to states in $\eta$ simply by time-passage; and
- every state in the time-passage also satisfies condition $\eta'$.

$\texttt{Tbck}(\eta', \eta)$ can be constructed as $\exists t(t \geq 0 \wedge \eta + t \wedge \texttt{path}(\eta', t))$ [14].

Given a partial assignment $\Pi$ of $G \cup L_1 \cup L_2 \cup X_1 \cup X_2$, we let $\eta \Pi$ be the precondition to $\eta$ before the assignment. Suppose $\Pi$ is defined for $\{y_1, \ldots, y_n\}$.

$$\eta \Pi \equiv (\bigwedge_{x \text{ is a clock defined in } \Pi.} x \geq 0) \wedge \exists y_1 \ldots \exists y_n (\eta \wedge \bigwedge_{1 \leq i \leq n} y_i = \Pi(y_i))$$

Given $e_1 \in E_1$ and $e_2 \in E_2^{(e_1)}$ such that $e_1$ and $e_2$ are compatible, the weakest precondition to $\eta$ through discrete transition pair $(e_1, e_2)$ can be represented as $\texttt{Xbck}_{(e_1, e_2)}(\eta) \equiv \tau_1(e_1) \wedge \tau_2(e_2) \wedge (\eta(\pi_1(e_1)\pi_2(e_2)))$.

With procedures $\texttt{Xbck}_{(e_1,e_2)}()$ and $\texttt{Tbck}()$, we can construct the symbolic backward reachability procedure, denoted $\texttt{Rbck}_{E_2^{(\perp)}}(\eta_1, \eta_2)$ for convenience, as in [14, 19]. Intuitively, $\texttt{Rbck}_{E_2^{(\perp)}}(\eta_1, \eta_2)$ characterizes the state-space for $\exists \eta_1 \mathcal{U} \eta_2$ through transitions in $E_2^{(\perp)}$. Computationally, $\texttt{Rbck}_{E_2^{(\perp)}}(\eta_1, \eta_2)$ is the least fixpoint solution of equation: $Y = \eta_2 \vee \texttt{Tbck}\left(\eta_1, \bigvee_{e_2 \in E_2^{(\perp)}} \texttt{Xbck}_{(\perp, e_2)}(Y)\right)$. That is, $\texttt{Rbck}_{E_2^{(\perp)}}(\eta_1, \eta_2) \equiv lfpY.\left(\eta_2 \vee \texttt{Tbck}\left(\eta_1, \bigvee_{e_2 \in E_2^{(\perp)}} \texttt{Xbck}_{(\perp, e_2)}(Y)\right)\right)$. Now we need to construct a symbolic characterization of $(\nu_1, \nu_2)$ for *stutter*$(\nu_1, \nu_2, \delta, \nu_2', \mathcal{Q})$ when $\nu_2'$ and $\mathcal{Q}$ are represented as formulas $(f_{\nu_2'})$ and $\mathcal{Q}$ respectively. We use an auxiliary clock $z \notin X_1 \cup X_2$ to measure the length of the stuttering run. Then, we have

$$f_{stutter}(\delta, f_{\nu_2'}, \mathcal{Q}) \equiv \exists z (z = 0 \wedge \texttt{Rbck}_{E_2^{(\perp)}}(\mathcal{Q}, z = \delta \wedge f_{\nu_2'}))$$

### 5.3 Construction of formula (B)

We rewrite formula (B) in formula (A) as follows.

$$\nu_1 + \delta \models \tau_1(e_1) \wedge (\nu_1 + \delta)\pi_1(e_1) \models H_1 \wedge \neg \exists 0 \leq \delta' \leq \delta(\nu_1 + \delta' \models \neg H_1) \ \ldots \ldots (B)$$

The first conjunct says that after $\delta$ time units from state $\nu_1$, $A_1$ satisfies the triggering condition of transition $e_1$. The second conjunct says that in this time-progression of $\delta$ time units, $A_1$ always satisfies $H_1$. The third conjunct says that at the end of the time-progression, $A_1$ goes from $\nu_1 + \delta$ to $(\nu_1 + \delta)\pi_1(e_1)$ and still satisfies $H_1$. The weakest characterization, i.e. formula (B), of $\nu_1$ can be derived backward from $H_1$ as follows.

$$(\text{Xbck}_{(e_1,\perp)}(H_1) + \delta) \wedge \text{path}(\neg H_1, \delta) \quad\dots\dots\dots\dots\dots\dots\dots\dots\dots\dots(f_B)$$

It characterizes those states that can take transition $e_1$ at $\delta$ time units away. The first conjunct corresponds to the first two conjuncts in formula (B) while the last one to the last one in (B).

### 5.4 Construction of formula (C)

We rewrite formula (C) as follows.

$$\exists \nu_2' \in V_{A_2} \left( \begin{array}{l} stutter_2(\nu_1, \nu_2, \delta, \nu_2', Q) \\ \wedge \exists e_2 \in E_2^{(e_1)} \left( \begin{array}{l} \nu_2' \xrightarrow{e_2} \nu_2' \pi_2(e_2) \\ \wedge \left( (\nu_1 + \delta)\pi_1(e_1), \nu_2'\pi_2(e_2) \right) \in Q \end{array} \right) \end{array} \right) \quad\dots\dots\dots(C)$$

The first conjunct in the outer quantification is for the measurement of the length, $\delta$, of the stuttering run through transitions internal to $A_2$. The second is for the execution of $e_2 \in E_2^{(e_1)}$. Specifically, the quantified $\nu_2'$ is for the precondition of $e_2$. Given a simulation relation representation $\mathcal{Q}$, the constraint of $\nu_2'$ from the second conjunct is as follows.

$$\bigvee\nolimits_{e_2 \in E_2^{(e_1)}} \text{Xbck}_{(e_1,e_2)}(\mathcal{Q}) \quad\dots\dots\dots\dots\dots\dots\dots\dots\dots\dots\dots(f_{\nu_2'})$$

To make sure that $A_1$ and $A_2$ observe the same behavior with $e_1$ and $e_2$ respectively, we construct the precondition of both $e_1$ and $e_2$ out of $\mathcal{Q}$ instead of the representation of $\nu_2$. Now with the formulations of $f_{\nu_2'}$ in the above and $f_{stutter}(\delta, f_{\nu_2'}, \mathcal{Q})$ in page 8, we find the following formulation for formula (C).

$$\exists z(z = 0 \wedge \text{Rbck}_{E_2^{(\perp)}}(\mathcal{Q}, z = \delta \wedge \bigvee\nolimits_{e_2 \in E_2^{(e_1)}} \text{Xbck}_{(e_1,e_2)}(\mathcal{Q}))) \quad\dots\dots\dots\dots(f_C)$$

### 5.5 Procedure

With formulas $(f_B)$ and $(f_C)$, we find that formula (A) can be constructed as

$$f_A(e_1, \mathcal{Q}) \equiv \exists \delta \geq 0 \left( \begin{array}{l} (\text{Xbck}_{(e_1,\perp)}(H_1) + \delta) \wedge \text{path}(\neg H_1, \delta) \\ \wedge \neg \exists z \left( z = 0 \wedge \text{Rbck}_{E_2^{(\perp)}} \left( \mathcal{Q}, z = \delta \wedge \bigvee\nolimits_{e_2 \in E_2^{(e_1)}} \text{Xbck}_{(e_1,e_2)}(\mathcal{Q}) \right) \right) \end{array} \right)$$

With formula $f_A(e_1, \mathcal{Q})$, we are now ready to present our procedure for simulation checking. The procedure is a greatest-fixpoint one. We start from $\mathcal{Q} = H_1 \wedge H_2$. Then we iteratively delete state pairs described in $f_A(e_1, \mathcal{Q})$ for each $e_1 \in E_1$ until a fixpoint is reached. For convenience, we let $FM\_elm(F, \{y_1, \dots, y_n\}) = \exists y_1 \exists y_2 \dots \exists y_n(F)$.

---

Simulation_Check$(A_1, A_2)$ /* $A_i = \langle \Sigma, X_i, G, L_i, I_i, H_i, E_i, \epsilon_i, \tau_i, \pi_i \rangle$, $1 \leq i \leq 2$ */ {
   let $\mathcal{Q} := H_1 \wedge H_2$; $\mathcal{Q}' := \textit{false}$; $\dots\dots\dots\dots\dots\dots\dots\dots\dots\dots\dots\dots\dots\dots\dots\dots$(F)
   while $(\mathcal{Q} \neq \mathcal{Q}')$, do {
     $\mathcal{Q}' := \mathcal{Q}$;
     for each $e_1 \in E_1$, $\mathcal{Q} := \mathcal{Q} \wedge \neg f_A(e_1, \mathcal{Q})$;
     if $(I_1 \neq FM\_elm(I_1 \wedge I_2 \wedge \mathcal{Q}, L_2 \cup X_2))$ $\dots\dots\dots\dots\dots\dots\dots\dots\dots\dots\dots\dots$(J)

    print "$A_1$ does not implement $A_2$." and return *false*.
  }
  print "$A_1$ implements $A_2$." and return *true*.
}

We can start the greatest fixpoint image from $H_1 \wedge H_2$ at statement (F) because of the following lemma.

**Lemma 1.** *If $A_1 \preceq A_2$, then there is a $Q$ such that $A_1 \preceq_Q A_2$ and $\forall (\nu_1, \nu_2) \in Q(\nu_1 \nu_2 \models H_1 \wedge H_2)$.* ∎

In statement (J), we check whether all initial states of $A_1$ is paired with some initial states of $A_2$ in $Q$. The statement employs a technique called *early decision of greatest fixpoint (EDGF)* [21] and could significantly reduce the computation time of greatest fixpoint evaluation when no simulation relation exists. The following lemma establishes the correctness of our procedure.

**Lemma 2.** *When Simulation_Check($A_1, A_2$) halts, it returns true iff $A_1 \preceq A_2$.* ∎

## 6   Algorithm with zone-technology

When the initial condition, invariance condition, and transition triggering conditions of $A_1$ and $A_2$ are all presented as LH-predicates, we can prove that all operations in Simulation_Check($A_1, A_2$) yield only LH-predicates. This is based on the following lemma and all our operations in Simulation_Check($A_1, A_2$) are based on the basic operations listed in the following lemma.

**Lemma 3.** *Given LH-predicates $\eta_1$ and $\eta_2$, $\neg \eta_1$, $\eta_1 \vee \eta_2$, $\eta_1 + \delta$, $\exists x(\eta_1)$ with $x \in X_1 \cup X_2$, and $\exists p(\eta_1)$ with $p \in G \cup L_1 \cup L_2$ are all LH-predicates.* ∎

However, the representation and manipulation of LH-predicates are usually less than efficient. In this section, we present techniques that allow us to implement procedure Simulaton_Check() with the zone-technology. The correctness of such techniques is based on theorem 10 in [18] which asserts that for any $\nu_1, \nu'_1$ in the same *region*[2] of $A_1$ and $\nu_2, \nu'_2$ in the same region of $A_2$, $(\nu_1, \nu_2) \in Q$ iff $(\nu'_1, \nu'_2) \in Q$. In the following, we carefully examine formula ($f_A$) for operations that yield non-zone-predicates and discuss how to rewrite the the predicates to make it representable with zones. We find two classes of operations that create non-zone-predicates. We present techniques in subsections 6.1 and 6.2 to represent the results of such operations with zones. In subsection 6.3, we combine the technques of the two subsections and reformulate procedure Simulation_Check() as an algorithm.

Formulas ($f_B$) and ($f_C$) both use basic procedures $\text{Rbck}_{E_2^{(\perp)}}()$, $\text{Tbck}()$, and $\text{Xbck}_{(e_1,e_2)}()$, which are in turn built upon Fourier-Motzkin elimination, Boolean operations, and "$+\delta$" operations of predicates. We let $ZP_c(X, P)$ be the set of all zone-predicates whose atoms are either atomic propositions in $P$ or inequalities like $x_1 - x_2 \sim d$ where $x_1, x_2 \in X \cup \{0\}$, '$\sim$'$\in \{<, \leq\}$, and $|d| \leq c$. According to the literature [14, 19], we find that given zone-predicates in $ZP_{C_{(A_1,A_2)}}(X, P)$ as their arguments, all such basic procedures yield zone-predicates in $ZP_{C_{(A_1,A_2)}}(X, P)$. The challenge here is that some arguments in formulas $(f_B)$ and $(f_C)$ are not exactly zone-predicates. Our idea is to rearrange the arguments to those basic procedures so that they all appear as zone-predicates in $ZP_{C_{(A_1,A_2)}}(X, G \cup L_1 \cup L_2)$ for some $X$.

---
[2] A *region* is a smallest state-space that can be characterized with a zone.

After examining formulas $(f_B)$ and $(f_C)$, we find out that there are only two ways that we may yield non-zone-predicates.

### 6.1 Time-progress operations in formulas $(f_B)$ and $(f_C)$

The first class of opertions happens when we execute the "$+\delta$" operation in formula $(f_B)$ and when we call the "$+t$" and "$+t'$" operations in procedures Tbck() and Rbck$_{E_2^{(\bot)}}$(). In these cases, $\delta, t, t'$ are not exactly clocks (i.e., their values do not change with time). Now we focus on the case of "$+\delta$." The other two cases for $t, t'$ are similar. After operation "$+\delta$," we convert literals like $x \sim c$ or $-x \sim -d$ respectively to something like $x + \delta \sim c$ and $-x - \delta \sim -d$ which do not look like atomic constraints in zone-predicates. What we do is that we introduce a new dense variable '$-\delta$' and instead convert those literals to $x - (-\delta) \sim c$ and $(-\delta) - x \sim -d$. In this way, given any argument in $ZP_{C_{(A_1,A_2)}}(X, P)$, the "$+\delta$" operations (and "$+t$" and "$+t'$" operations) all yield zone-predicates in $ZP_{C_{(A_1,A_2)}}(X \cup \{-\delta, -t, -t'\}, P)$. So we can establish the following lemma.

**Lemma 4.** *Given TEAs $A_1$ and $A_2$ and $\eta \in \text{ZP}_{C_{(A_1,A_2)}}(G \cup L_1 \cup L_2, X_1 \cup X_2)$, $\eta + \delta$ and* path$(\neg\eta, \delta)$ *are both in* $\text{ZP}_{C_{(A_1,A_2)}}(G \cup L_1 \cup L_2, X_1 \cup X_2 \cup \{-\delta\})$. ∎

Note that the correctness of this first conversion relies on the fact that we never do a double time-progression operation like $(\eta + t) + t'$ in $(f_B)$ and $(f_C)$.

### 6.2 Measuring time-progress with $\delta$ and clock $z$ in formula $(f_C)$

The second way that we may yield non-zone-predicate stems from equality $z = \delta$ in formula $(f_C)$. This equality is represented as zone-predicate $z - \delta \leq 0 \wedge \delta - z \leq 0$. This could make trouble since when we apply the "$+t$" (or "$+t'$") operations in formula $(f_C)$, the literals are converted to $z + t - \delta \leq 0$ and $\delta - z - t \leq 0$ since $\delta$ does not change its values with time progress. Such literals are certainly not zone-predicates. One observation from formula $(f_A)$ is that the quantification of $z$ appears inside that of $\delta$. Thus in the scope of processing $z$-related predicates in formula $(f_C)$, $\delta$ is static and stays unchanged. Moreover, only the "$+t$" and "$+t'$" operations appear in formula $(f_C)$ while no "$+\delta$" operation does. Thus our idea is to use auxiliary clock '$z - \delta$' instead of $z$. Note that clock '$z - \delta$' is special in that its value may be less than zero. Then equality $z = \delta$ is instead represented as $(z-\delta) \leq 0 \wedge -(z-\delta) \leq 0$ and $(z = \delta) + t$ yields $(z-\delta)-(-t) \leq 0 \wedge (-t)-(z-\delta) \leq 0$, which again falls in the syntax of zone-predicates.

There is one technicality that we need to take care of after introducing auxiliary clock '$z - \delta$.' In formula $(f_C)$, we need to evaluate

$$\exists z(z = 0 \wedge \texttt{Rbck}_{E_2^{(\bot)}}(\mathcal{Q}, z = \delta \wedge \bigvee_{e_2 \in E_2^{(e_1)}} \texttt{Xbck}_{(e_1,e_2)}(\mathcal{Q}))) \dots\dots\dots\dots (K)$$

With the explanation in the previous paragraph, we can prove the following lemma.

**Lemma 5.** *Given TEAs $A_1$ and $A_2$ and $\eta \in \text{ZP}_{C_{(A_1,A_2)}}(G \cup L_1 \cup L_2, X_1 \cup X_2)$, $\texttt{Rbck}_{E_2^{(\bot)}}(\eta, z = \delta \wedge \bigvee_{e_2 \in E_2^{(e_1)}} \texttt{Xbck}_{(e_1,e_2)}(\eta))$ is in $\text{ZP}_{C_{(A_1,A_2)}}(X_1 \cup X_2 \cup \{(z-\delta)\}, G \cup L_1 \cup L_2)$.* ∎

With lemma 5, we can assume that formula $\texttt{Rbck}_{E_2^{(\bot)}}(\mathcal{Q}, z = \delta \wedge \bigvee_{e_2 \in E_2^{(e_1)}} \texttt{Xbck}_{(e_1,e_2)}(\mathcal{Q}))$ yields a zone-predicate $f_M$. Formula $(K)$ can be evaluated as zone-predicates by replacing every occurrence of $z$ in $f_M$ with 0. We can implement a procedure, $\texttt{replace}(\eta, z)$, that replace every occurrence of clock variable '$z - \delta$' in zone-predicate $\eta$ with clock variable value '$-\delta$.' Thus $\exists(z = 0 \wedge f_M) \equiv \texttt{replace}(f_M, z)$.

### 6.3 Implementing Simulation_Check() with zones

Combining the techniques in the previous two subsections, we can establish the following three lemmas whose proof are omitted due to page-limit.

**Lemma 6.** *When $A_1$ and $A_2$ are both TEAs, $f_A(e_1, \mathcal{Q})$ is equivalent to*

$$\exists \delta \geq 0 \begin{pmatrix} \texttt{Xbck}_{(e_1,\bot)}(H_1) + \delta \wedge \texttt{path}(\neg H_1, \delta) \\ \wedge \neg \texttt{replace}(\texttt{Rbck}_{E_2^{(\bot)}}(\mathcal{Q}, z - \delta = 0 \wedge \bigvee_{e_2 \in E_2^{(e_1)}} \texttt{Xbck}_{(e_1,e_2)}(\mathcal{Q})), z, 0) \end{pmatrix}$$
∎

With lemmas 6, 4, and 5, the main result of this section is established as follows.

**Theorem 1.** *Simulation_Check($A_1, A_2$) is implementable as an algorithm with zones.* ∎

## 7 Algorithm for NZ-simulation checking

In this section, we present the following lemma that helps us adapting procedure Simulation_Check() for the checking of NZ-simulation. Please be reminded that in definition 5, $NZ_1$ denotes a representation of the non-Zeno states in the reachable state-space of $A_1$. The construction of zone-predicates for $NZ_1$ was discussed in [14, 21].

**Lemma 7.** *Given $A_i = \langle \Sigma, X_i, G, L_i, I_i, E_i, \epsilon_i, \tau_i, \pi_i \rangle$, $1 \leq i \leq 2$, if $A_1 \preceq^{NZ} A_2$, then $A_1 \preceq_Q^{NZ} A_2$ for some $Q$ such that for all $(\nu_1, \nu_2) \in Q$, $\nu_1 \models NZ_1$.*
**Proof :** We assume there is $(\nu_1, \nu_2) \in Q$ such that $\nu_1$ is a Zeno state $\in V_{A_1}$. According to definition 5, by deleting all such pairs from $Q$, we still get an NZ-simulation relation out of $Q$. Thus the lemma is proven. ∎

Lemma 7 leads to the following algorithm for NZ-simulation-checking.

---

NZ-Simulation_Check($A_1, A_2$)
/* $A_i = \langle \Sigma, X_i, G, L_i, I_i, H_i, E_i, \epsilon_i, \tau_i, \pi_i \rangle$, $1 \leq i \leq 2$ */ {
    Construct $NZ_1$ and let $A_1'$ be $\langle \Sigma, X_1, G, L_1, I_1 \wedge NZ_1, NZ_1, E_1, \epsilon_1, \tau_1, \pi_1 \rangle$.
    Return Simulation_Check($A_1', A_2$).
}

---

## 8 Experiments

We have implemented the techniques discussed in this manuscript in **RED** 7.0, a model-checker for TEAs and parametric safety analysis for LHAs based on CRD and HRD-technology [19, 20]. We have experimented with the following parameterized benchmarks with various numbers of processes. $A_1$ and $A_2$ differ in only one process.

- *Fischer's timed mutual exclusion algorithm* [5]: The algorithm relies on a global lock and a local clock per process to control access to the critical section. Two timing constants used are 10 and 19. We use two versions of this benchmark, one with a simulation relation and one without.
- *CSMA/CD benchmark* [22]: This is the ethernet bus arbitration protocol with the idea of collision-and-retry. The timing constants used are 26, 52, and 808. We use three versions of this benchmark, one with an NZ-simulation relation, one with a simulation relation, and one without.
- *Timed consumer/producer*: There are a buffer, some producers, and some consumers in the benchmark. The producers periodically write data to the buffer. The consumers also periodically wipe out data, if any, in the buffer. We use two versions of this benchmark, one with the biggest timing constant 15 and a simulation relation while the other with the biggest timing constant 20 and without a simulation relation.

The performance data is reported in table 1. For each row, we report the computation time for constructing $NZ_1$ and the time for simulation-checking. The total memory consumption for the data-structures in state-space reprsentations is also reported. In this experiment, we did not run benchmarks with large concurrency sizes. But according to the grow-rates of the memory consumptions, we predict that benchmarks with larger concurrency sizes could be passed with our program.

## 9 Conclusion

In this work, we present a characterization of the simulation relation between TEAs and derive a symbolic simulation-checking procedure out of this characterization. We then present techniques to implement the algorithm with zone-technology. It would be interesting to see what classes of LHAs can be verified with zone-technology using data collected on a Pentium 4 1.7GHz with 380MB memory running LINUX;
s: seconds; k: kilobytes of memory in data-structure;

**Table 1.** Performance data of scalability w.r.t. various strategies

| benchmarks | versions | $m$ | time non-Zeno restriction | time Simulation Check | total time and memory |
|---|---|---|---|---|---|
| Fischer's mutual exclusion ($m$ processes) | Simulation exists. | 1 | 0.01s | 0.00s | 0.01s/23k |
| | | 2 | 0.31s | 0.43s | 0.74s/90k |
| | | 3 | 1.27s | 2.30s | 3.57s/201k |
| | | 4 | 4.24s | 8.27s | 12.51s/431k |
| | | 5 | 12.98s | 26.84s | 39.82s/897k |
| | Simulation does not exist. | 1 | 0.01s | 0.00s | 0.01s/22k |
| | | 2 | 0.27s | 0.11s | 0.38s/88k |
| | | 3 | 1.42s | 0.65s | 1.73s/190k |
| | | 4 | 3.53s | 1.93s | 5.46s/390k |
| | | 5 | 10.62s | 7.26s | 17.90s/792k |
| CSMA/CD (1 bus+ $m$ senders) | Simulation exists. | 1 | 0.02s | 0.00s | 0.02s/44k |
| | | 2 | 0.25s | 0.36s | 0.61s/161k |
| | | 3 | 2.15s | 88.09s | 90.24s/3681k |
| | Only NZ-simulation exists. | 1 | 0.18s | 0.03s | 0.21s/53k |
| | | 2 | 1.12s | 2.10s | 2.73s/199k |
| | | 3 | 5.90s | 122.0s | 127.9s/2447k |
| | No simulation exists. | 1 | 0.03s | 0.01s | 0.04s/45k |
| | | 2 | 0.26s | 0.90s | 1.16s/183k |
| | | 3 | 2.28s | 25.82s | 28.10s/4365k |
| Consumer & producer (1 buffer +1 producer +$m$ consumers) | Simulation exists. | 1 | 0.07s | 0.00s | 0.07s/39k |
| | | 2 | 0.24s | 0.03s | 0.27s/48k |
| | | 3 | 0.62s | 0.05s | 0.67s/76k |
| | | 4 | 2.01s | 0.08s | 2.09s/173k |
| | | 5 | 6.51s | 0.21s | 6.72s/403k |
| | does not exist. | 1 | 0.06s | 0.03s | 0.09s/52k |
| | | 2 | 0.28s | 0.23s | 0.51s/61k |
| | | 3 | 0.70s | 0.22s | 0.92s/104k |
| | | 4 | 2.75s | 0.33s | 3.08s/245k |
| | | 5 | 10.64s | 0.92s | 11.56s/590k |

this technique. Our algorithm can also be adapted to handle the effect of Zeno states. Finally, our implementation and experiment shows the promise that our algorithm could be useful in practice in the future.

# APPENDIX

## A  Definition of non-Zeno simulation and equivalence

Suppose we are given two TEAs $A_1$ and $A_2$ such that $A_i = \langle \Sigma, X_i, G, L_i, I_i, H_i, E_i, \epsilon_i, \tau_i, \pi_i \rangle$, $1 \leq i \leq 2$, and two states $\nu_1 \in V_{A_1}$ and $\nu_2 \in V_{A_2}$. Let $NZ_1$ be a representation of the non-Zeno states in the reachable state-space of $A_1$. An *NZ-simulation* $Q$ from $A_1$ to $A_2$ is a binary relation from $V_{A_1}$ to $V_{A_2}$ that satisfies the following requirement. For every $(\nu_1, \nu_2) \in Q$, $\nu_1$ and $\nu_2$ agrees on interpretation of variables in $G$ and for every $\delta \in \mathbb{R}^{\geq 0}$ and transition $e_1$ of $E_1$ such that for all $\hat{\delta} \in [0, \delta]$, $\nu_1 + \hat{\delta} \models H_1$ and $\nu_1 + \delta \xrightarrow{e_1} \nu_1'$, there are $\nu_2' \in V_{A_2}$ and $e_2 \in E_2^{(e_1)}$ such that $stutter_2(\nu_1, \nu_2, \delta, \bar{\nu}_2, Q)$, $\nu_2' \xrightarrow{e_2} \nu_2' \pi_2(e_2)$, and $(\nu_1', \nu_2' \pi_2(e_2)) \in Q$. Or in logic notations,

$$\forall (\nu_1, \nu_2) \in Q \forall e_1 \in E_1 \forall \delta \in \mathbb{R}^{\geq 0}$$
$$\left( \begin{pmatrix} \nu_1 + \delta \models \tau_1(e_1) \\ \land \forall 0 \leq \delta' \leq \delta(\nu_1 + \delta' \models H_1) \\ \land (\nu_1 + \delta)\pi_1(e_1) \models emnz_1 \land H_1 \end{pmatrix} \longrightarrow \exists \nu_2' \in V_{A_2} \begin{pmatrix} stutter(\nu_1, \nu_2, \delta, \nu_2', Q) \\ \land \exists e_2 \in E_2^{(e_1)} \begin{pmatrix} \nu_2' \xrightarrow{e_2} \nu_2' \pi_2(e_2) \\ \land ((\nu_1 + \delta)\pi_1(e_1), \nu_2' \pi_2(e_2)) \in Q \end{pmatrix} \end{pmatrix} \right)$$

If there is an NZ-simulation relation $Q$ from $A_1$ to $A_2$ such that for every state $\nu_1 \models I_1 \land NZ_1$, there is a $\nu_2 \models I_2$ such that $(\nu_1, \nu_2) \in Q$, we denote $A_1 \preceq_Q A_2$. If $\exists Q(A_1 \preceq_Q A_2)$, we say $A_1$ *implements* (or *is simulated by*) $A_2$, in symbols $A_1 \preceq A_2$. Two TEAs $A_1$ and $A_2$ *are equivalent* (i.e. *bisimulate*), in symbols $A_1 \equiv A_2$, if $A_1 \preceq A_2$ and $A_2 \preceq A_1$. ∎

## B  Fourier-Motzkin elimination for the special case

There are two cases to discuss. The first case is for calculating $\exists p_1(f(p_1, \ldots, p_m, x_1, \ldots, x_n))$. According to Shannon expansion, we have

$$\exists p_1(f(p_1, \ldots, p_m, x_1, \ldots, x_n)) \equiv$$
$$f(false, \ldots, p_m, x_1, \ldots, x_n) \lor f(true, \ldots, p_m, x_1, \ldots, x_n)$$

The second case, for the calculation of $\exists x_1(f(p_1, \ldots, p_m, x_1, \ldots, x_n))$, can be handled with the following steps.
(1) Rewriting $f(p_1, \ldots, p_m, x_1, \ldots, x_n)$ in disjunctive normal form.
(2) For each disjunct,
    (2.1) rewrite the linear constraints in one of the following three forms.
        TYPE I: $\sum_{2 \leq i \leq n} a_i x_i \sim -a_1 x_1 + d$ when $a_1 > 0$.
        TYPE II: $\sum_{2 \leq i \leq n} b_i x_i \sim' -b_1 x_1 - d'$ when $b_1 < 0$.
        TYPE III: $\sum_{2 \leq i \leq n} c_i x_i \sim d$ when $c_1 = 0$.
    (2.2) for each TYPE I constraint $\sum_{2 \leq i \leq n} a_i x_i \sim -a_1 x_1 + d$ and TYPE II constraint $\sum_{2 \leq i \leq n} b_i x_i \sim' -b_1 x_1 - d'$, conjunct the following constraint to the disjunct.

$$\sum_{2\leq i\leq n}(a_i|b_1| + b_i|a_1|)x_i \sim'' |b_1|d + |a_1|d'$$
where $\sim''=$ '$\leq$' if both $\sim=$ '$\leq$' and $\sim=$ '$\leq$'; and $\sim''=$ '$<$' otherwise.

(2.3) delete every constraint with a non-zero coefficient for $x_1$ in the disjunct.